\documentclass{aa520}

\usepackage{graphicx}

\def\sbs{SBS\,1150+599A}
\def\png{PN\,G\,135.9+55.9}

\newcommand{\Ms}{~M$_{\odot}$}

\newcommand{\qh}{$Q({\rm{H^{0}}})$}

\newcommand{\Tstar}{$T_{\star}$}
\newcommand{\Lstar}{$L_{\star}$}

\newcommand{\Ha}{H$\alpha$}
\newcommand{\Hb}{\ifmmode {\rm H}\beta \else H$\beta$\fi}

\newcommand{\Hei}{He~{\sc i} $\lambda$5876}

\newcommand{\Nii}{[N~{\sc ii}] $\lambda$6584}

\newcommand{\oii}{[O~{\sc ii}]}
\newcommand{\Oiii}{[O~{\sc iii}] $\lambda$5007}

\newcommand{\oiii}{[O~{\sc iii}]}
\newcommand{\Neiii}{[Ne~{\sc iii}] $\lambda$3869}
\newcommand{\neiii}{[Ne~{\sc iii}]}
\newcommand{\nev}{[Ne~{\sc v}]}
\newcommand{\Nev}{[Ne~{\sc v}] $\lambda$3426}
\newcommand{\Sii}{[S~{\sc ii}] $\lambda$6716, $\lambda$6731}

\newcommand{\Ariii}{[Ar~{\sc iii}] $\lambda$7135}

\newcommand{\Ho}{H$^{0}$}
\newcommand{\Hp}{H$^{+}$}

\newcommand{\Hep}{He$^{+}$}
\newcommand{\Hepp}{He$^{++}$}

\newcommand{\Opp}{O$^{++}$}

\newcommand{\Nepp}{Ne$^{++}$}

\newcommand{\Arppp}{Ar$^{+++}$}

\newcommand{\Siii}{[S~{\sc iii}] $\lambda$9069}

\begin{document}

         \title{The $\alpha$-element abundances in the most oxygen-poor
         planetary nebula PN G 135.9+55.9}

       \subtitle{}
       \titlerunning{The $\alpha$-element abundances in \png}

       \author{Richer, M. G.
               \inst{1}
               \thanks{Visiting Astronomer, Observatorio
               Astr\'onomico Nacional in San Pedro M\'artir, Mexico.}
               \and
               Tovmassian, G.
               \inst{1} \and
               Stasi\'nska, G.
               \inst{2} \and
               Jameson, R. F.
               \inst{3} $^{\star\star}$ \and
               Dobbie, P. D.
               \inst{3}
               \thanks{Visiting Astronomer,
               William Herschel Telescope, operated by the Isaac Newton
               Group (ING) on behalf of the
               UK Particle Physics and Astronomy Research Council
               (PPARC), the Nederlanse Organisatie voor Wetenschappelijk
               Onderzoek (NWO), and the Instituto de Astrofísica de
               Canarias (IAC).}
               \and
               Veillet, C.
               \inst{4}
               \thanks{Visiting Astronomer, Canada-France-Hawaii Telescope
               operated by the National Research Council of Canada, the
               Centre National de la Recherche Scientifique de France and
               the University of Hawai'i.}
               \and
               Gutierrez, C.
               \inst{5}
               \thanks{Visiting Astronomer,
               Nordic Optical Telescope, operated by the ING on
               behalf of the PPARC, NWO, and IAC.}
               \and
               Prada, F.
               \inst{6} $^\dagger$
       }

       \offprints{Michael Richer}

       \institute{Observatorio Astron\'omico Nacional, Instituto de
                  Astronom\'\i a, UNAM, P.O. Box 439027, San Diego,
                  CA, USA 92143-9027 \\
                  \email{\{richer, gag\}@astrosen.unam.mx}
           \and
                  LUTH, Observatoire de Meudon, 5 Place Jules Janssen,
                  F-92195 Meudon Cedex, France \\
                  \email{grazyna.stasinska@obspm.fr}
           \and
                  University of Leicester, University Road, Leicester
                  LE1 7RH, UK \\
                  \email{\{rfj, pdd\}@star.le.ac.uk}
           \and
                  Canada-France-Hawaii Telescope Corp., P.O. Box 1597,
                  Kamuela, HI, USA 96743 \\
                  \email{veillet@cfht.hawaii.edu}
           \and
                  Instituto de Astrof\'\i sica de Canarias, C/V\'\i a
                  L\'actea s/n, E-38200, La Laguna, Tenerife, Spain \\
                  \email{cgc@ll.iac.es}
           \and
                  Centro Astron\'omico Hispano Alem\'an, C/Jes\'us
                  Durb\'an Rem\'on 2-2, E-04004, Almer\'\i a,
                  Spain\\ \email{prada@caha.es}
       }

       \date{Received ???; accepted ???}

       \abstract{
       We present extensive new spectroscopy and imaging of \png.
       We use these data as constraints to photoionization models
       to derive limits on the oxygen abundance.
       We find that \png\ has an oxygen abundance less than 1/50 of
       the solar value.
       Our models favour a value of $12 + \log \mathrm O/\mathrm H$
       between 5.8 and 6.5\,dex,
       confirming that \png\ is the most
       oxygen-poor planetary nebula known (Tovmassian et al.
       \protect\cite{tovmassianetal2001}). We
       also derive $\mathrm{Ne}/\mathrm O = 0.5 \pm 0.3$, $\mathrm
       S/\mathrm O < 0.094$, and $\mathrm{Ar}/\mathrm O < 0.23$.
       Although the value of Ne/O is nominally high, it need not
       imply that
       the progenitor of \png\ converted any of its initial oxygen
       abundance to neon.
       The helium abundance  appears to be very low, $\mathrm{He}/\mathrm
       H\sim 0.08$, but a precise determination will require a much
       more detailed study.  We find that $\mathrm H\alpha/\mathrm
       H\beta$ is lower than expected and
       perhaps variable, a finding for which we have no clear explanation.
       }

      \maketitle

%

\begin{table*}
\caption[]{Log of the spectroscopic observations}
\label{speclog}
\[
\begin{tabular}{lcccccc}
\hline \noalign{\smallskip}
   & SPM1 & SPM2 & CFHT & WHT & SPM3$^{\mathrm a}$ & SPM4 \\
\noalign{\smallskip}\hline\noalign{\smallskip}
date & 22 Jan 2001 & 23-24 Jan 2001 & 3-4 Mar 2001 & 11 May 2001 & 5
Mar 2002 & 8 Apr 2002\\
CCD  & Thomson  & Thomson     & EEV        & Tektronix & SITe & SITe
\\
format & $2048\times 2048$  & $2048\times 2048$ & $2048\times 4500$ &
$1024\times 1024$ & $1024\times 1024$ & $1024 \times 1024$\\
pixel size$^\mathrm b$ & $14\,\mu$m, $0\farcs61$ & $14\,\mu$m,
$0\farcs61$ & $13.5\,\mu$m, $0\farcs283$ & $24\,\mu$m, $0\farcs36$ &
$24\,\mu$m, $1\farcs05$ & $24\,\mu$m, $1\farcs05$ \\
gain (e$^-$/ADU) & 0.5 & 0.5 & 1.8 & 1.4 & 1.3 & 1.3\\
read noise (e$^-$) & 4.8 & 4.8 & 3.1 & 4.6 & 8 & 8\\
spectrograph & B\&C & B\&C & MOS & ISIS & B\&C & B\&C\\
grating$^{\mathrm c}$ & 300/4550\AA & 600/4550\AA & B400/5186\AA &
R158R/6500\AA & 300/4550\AA & 400/5150\AA\\
slit width & 3\farcs8 & 3\farcs8 & 5\arcsec & 1\arcsec &2\farcs3 &
2\farcs3\\
spectral resolution$^{\mathrm d}$ & 12.7\AA & 5.6\AA & 23.0\AA &
7.8\AA & 8.1\AA & 5.8\AA \\
wavelength interval & 3700-6700\AA & 4745-7200\AA & 3400-8000\AA &
6820-9740\AA & 3600-7100\AA & 3660-6760\AA \\
arc lamp & HeAr & HeAr & HgNeAr & CuNe+CuAr & HeAr & HeAr\\
standard stars & HD93521 & G191B2B & Feige 66 & Hz44 &
BD+33$^\circ$2642 & HD93521\\
                  & BD+33$^\circ$2642 & HD93521 & & & &
BD+33$^\circ$2642\\
total exp. time & 4500\,s & 7800\,s & 7200\,s & 3600\,s & 5400\,s &
12600\,s\\
number of spectra & 3 & 4 & 4 & 3 & 3 & 7 \\
\noalign{\smallskip}\hline
\end{tabular}
\]
\begin{list}{}{}
\item[$^{\mathrm a}$] These observations were obtained through
clouds.
\item[$^\mathrm b$] Both the physical pixel size and the angle
subtended on the sky are given.
\item[$^\mathrm c$] The gratings are described by their ruling
(lines/mm)
and effective blaze wavelength.  For the grism used at CFHT, the
ruling and zero deviation wavelength are given.  All of the
observations were obtained in first order.
\item[$^{\mathrm d}$] This is the spectral resolution measured at
H$\alpha$, defined as the FWHM of the H$\alpha$ line, except for
the WHT spectrum, where this is the FWHM of P9$\lambda$9229.
\end{list}
\end{table*}

\section{Introduction}

Recently, \sbs\ has been recognized as a planetary nebula in the
Galactic halo  by  Tovmassian et al.
(\cite{tovmassianetal2001}) and renamed \png. The spectra then
available for this object were quite unusual for a planetary
nebula, presenting only the Balmer lines of hydrogen, He~{\sc ii}
$\lambda\lambda$4686,5411, and very weak \Oiii\ ($\sim4 \%$ of
\Hb). A photoionization model analysis showed that such a spectrum
implies a strongly density bounded and extremely oxygen-poor
nebula ionized by a very hot star. The oxygen abundance was
estimated to be less than 1/50 of the solar value, and probably
between 1/100 and 1/500 of solar assuming canonical properties for
the central star, making of \png\ by far the most oxygen-poor
planetary nebula known, with an oxygen abundance similar to the
lowest measured to date in stars (Boesgaard et al
\cite{boesgaardetal1999}; Howard et al \cite{howardetal1997}).

In this paper, we report detailed follow-up observations, aimed at
providing more stringent constraints on the nature of this
exceptional object. Section 2 presents the new spectroscopic data,
while Section 3 deals with narrow-band imaging. In Section 4, we
present an updated photoionization model analysis, taking full
advantage of the constraints provided by our new observational
data. This leads to a limit on the oxygen abundance which is now
\emph{independent of any assumption about the evolutionary status
of the central star}. In Section 5, we estimate the abundances of
the other elements. Section 6 presents a brief concluding
discussion.

\section{Spectroscopy}

\subsection{Observations}

\begin{table*}
\caption[]{Raw line intensities relative to H$\beta$ for the
optical spectra$^\mathrm a$} \label{lineint}
\[
\begin{tabular}{lccccccc}
\hline \noalign{\smallskip} ion            & $\lambda$ & SPM1
& SPM2             & CFHT            & SPM3$^{\mathrm b}$ & SPM4
\\ \noalign{\smallskip}\hline\noalign{\smallskip} {[}\ion{Ne}{v}{}
& 3426 &$               $&$                $&$
< 9.6      $&               &                \\
{[}\ion{O}{ii}{]}   & 3727 &$               $&$                $&$
< 2.1      $&               &                \\
{[}\ion{Ne}{iii}{]} & 3869 &$               $&$                $&$
1.04\pm 0.52 $&               &                \\ \ion{H}{i}
& 3889 &$               $&$                $&$ 2.92\pm 0.73 $&
&                \\ \ion{H}{i}          & 3970 &$  10.0\pm 2.5
$&$                $&$ 6.11\pm 0.64 $& $4.9\pm 2.9$  &$4.5\pm 2.1$
\\ \ion{H}{i}          & 4101 &$  21.4\pm 2.8  $&$
$&$ 20.6\pm 1.2  $& $19.3\pm 3.7$ &$21.3\pm 2.8$   \\ \ion{C}{ii}
& 4267 &$               $&$                $&$
< 1.1      $&               &$< 3.0$         \\
\ion{H}{i}          & 4340 &$  41.0\pm 1.6  $&$                $&$
42.05\pm 0.98 $& $43.8\pm 2.4$ &$40.7\pm 1.4$   \\
{[}\ion{O}{iii}{]}  & 4363 &$ < 1.9$         &
&$< 0.88$         &               &$< 1.8$         \\ \ion{He}{ii}
& 4686 &$  76.1\pm 2.3  $&$ $&$78.6\pm 1.5    $& $73.0\pm 2.1$
&$77.8\pm 1.3$   \\ {[}\ion{Ar}{iv}{]}$^{\mathrm c}$& 4711 &$<
2.0$&                &$ < 1.0$         &               &$< 1.4$
\\ \ion{H}{i}          & 4861 &$ 100.0\pm 2.1  $&$ 100.00\pm 0.84
$&$ 100.0\pm 1.6  $&$100.0\pm 1.7$ &$100.00\pm 0.85$\\
{[}\ion{O}{iii}{]}  & 5007 &$   2.7\pm 1.3  $&$   2.17\pm 0.61 $&$
2.87\pm 0.85 $& $3.8\pm 1.1$  &$2.31\pm 0.41$  \\ \ion{He}{ii}
& 5412 &$  5.68\pm 0.72 $&$ 5.47\pm 0.62   $&$ 5.23\pm 0.34  $&
$5.64\pm 0.91$&$5.53\pm 0.46$  \\ \ion{He}{i}         & 5876 &$
< 0.47     $&$     < 0.69     $&$
< 0.28     $& $< 1.4$       &$< 0.62$        \\
\ion{H}{i}          & 6563 &$ 314.7\pm 7.6  $&$  296.8\pm 2.4  $&$
260.3\pm 4.4  $& $288.0\pm 4.1$&$276.6\pm 2.0$  \\
{[}\ion{N}{ii}{]}   & 6583 &    $< 3.2      $&$     < 1.4      $&$
< 0.45     $& $< 1.3$       &$< 0.74$        \\
{[}\ion{S}{ii}{]}   & 6716 &                 & &                 &
$< 1.3$       &$< 0.74$        \\ {[}\ion{S}{ii}{]}   & 6731 &
& &                 & $< 1.3$       &$< 0.74$        \\ $F(\mathrm
H\beta)^\mathrm d$ &&$ 1.82\pm 0.03$&$ 1.33\pm 0.01 $&$ 2.55\pm
0.03  $& $0.99\pm 0.01$&$1.14\pm 0.01$  \\ $W(\mathrm
H\beta)^\mathrm e$ &&$68.4\pm 1.4$& $61.4\pm 0.7$ &$65.3\pm 1.0$
&$61.5\pm 1.3$  &$53.5\pm 0.5$   \\ $W(\mathrm H\alpha)^\mathrm
e$&&$646\pm 58$  & $508\pm 24$ &$541\pm 32$      &$494\pm 10$
&$514\pm 10$     \\ H$\alpha$ range$^\mathrm f$ && 304--318      &
256--350         & 258--267        & 265--324      & 241--321 \\
\noalign{\smallskip}\hline
\end{tabular}
\]
\begin{list}{}{}
\item[$^{\mathrm a}$] When no uncertainty is given, the value
represents a $2\sigma$ upper limit to the flux in the line.
\item[$^{\mathrm b}$] These observations were obtained through
clouds.
\item[$^{\mathrm c}$] These upper limits also apply to
[\ion{Ar}{iv}]$\lambda$4740.
\item[$^{\mathrm d}$] This is the total flux in emission measured at
H$\beta$ in units of $10^{-14}$\,erg\,s$^{-1}$\,cm$^{-2}$.
\item[$^{\mathrm e}$] These are the equivalent widths in \AA\ of
H$\beta$ and H$\alpha$ in emission.
\item[$^{\mathrm f}$] This is the total range spanned by the
values of ${\mathrm H}\alpha/{\mathrm H}\beta$ among the
individual spectra.

\end{list}
\end{table*}

Table \ref{speclog} presents a summary of our new spectroscopic
observations of \png.  This table includes the dates of the
observations, the instrumental configuration, and the flux and
wavelength standards that were used.  The only observing run that
suffered from non-photometric conditions was that of 5 Mar 2002,
when significant cloud cover affected observations of both \png\
and the standard star.

The spectroscopy from the Observatorio Astron\'omico Nacional in
San Pedro M\'artir, Baja California, Mexico (SPM) was obtained
using the Boller \& Chivens spectrograph (B\&C) and three
different gratings during four observing runs. For the 2001
observations, a rather wide slit (3\farcs8) was used to better
measure the total fluxes, while, for the 2002 observations, a
narrower slit was used to obtain higher spectral resolution and
better sensitivity to fainter lines. The standard stars were
observed with an even wider slit (9\arcsec). In all cases, the
slit was oriented east-west on the sky.  Spectra of the
illuminated dome wall were obtained to serve as flat field images.
Bias images were obtained at the beginning and end of the night.

The spectroscopy at the Canada-France-Hawaii Telescope (CFHT) was
obtained with the Multi-Object Spectrograph (MOS; Le F\`evre et
al. \cite{lefevre1994}).  Both the object and the standard star
were observed through a 5\arcsec\ slit. However, the observations
of PN~G 135.9+55.9 were obtained the night before those of the
standard star.  Spectra of the internal halogen lamp were obtained
to serve as flat field images.

The spectroscopy at the William Herschel Telescope (WHT) was
obtained using the red arm of the ISIS spectrograph.  The object
was observed through a 1\arcsec\ slit, while the standard stars
were observed with a 10\arcsec\ slit.  For these observations, the
slit was oriented at the parallactic angle. Spectra of the
internal lamp were obtained to serve as flat field images while
spectra of the sky were used to correct for the slit illumination.

All of the spectroscopy  was reduced using the Image Reduction and
Analysis Facility (IRAF)\footnote{IRAF is distributed by the
National Optical Astronomical Observatories, which is operated by
the Associated Universities for Research in Astronomy, Inc., under
contract to the National Science Foundation.} software package
(specifically the specred package).  In all cases, the overscan
bias was subtracted from each image.  For the SPM data, the
overscan-subtracted bias images obtained during the night were
combined and subtracted from all of the images. Next, the
pixel-to-pixel variations were removed by division of the flat
field image.  For the WHT data, the slit illumination correction
was then applied.  The sky emission was subtracted during the
extraction of the one dimensional spectra by defining sky regions
on both sides of the object spectra and interpolating between them
with a straight line. The wavelength calibration was performed
using arc lamp spectra obtained at the time of the object
observations. Finally, the spectra were calibrated in flux using
the observations of the standard stars (Table \ref{speclog}) to
determine the instrumental sensitivity function.  The individual
spectra were calibrated in both wavelength and flux before being
summed together.

Table \ref{lineint} presents the raw line intensities 
relative to H$\beta$ measured in the optical spectral region for
PN~G 135.9+55.9.  The line intensities presented in Table
\ref{lineint} are those for the summed spectra from each observing
run.  The line intensities were measured using the software
described by McCall et al. (\cite{mccalletal1985}). The
uncertainties quoted for each line intensity ($1 \sigma$) include
contributions from the fit to the line itself, from the fit to the
reference line, and from the noise in the continuum for both the
line and reference line.  When only a limit is given, this
corresponds to a $2 \sigma$ upper limit to undetected lines.

\begin{table}
\caption[]{Raw line fluxes for the WHT spectra} \label{whtlineint}
\[
\begin{tabular}{lccccccc}
\hline \noalign{\smallskip} ion & $\lambda$ & flux$^\mathrm a$ &
$F(\lambda)/F(\mathrm H\beta)^\mathrm b$\\
\noalign{\smallskip}\hline\noalign{\smallskip} {[}\ion{Ar}{iii}{]}
& 7135 & $< 0.45$      & $< 0.06$       \\ \ion{H}{i}          &
8750 & $5.7\pm 2.3$  & $0.74\pm 0.31$ \\ {[}\ion{S}{iii}{]}  &
9069 & $< 0.59$      & $< 0.08$       \\ \ion{H}{i}          &
9229 & $20.3\pm 3.4$ & $2.64\pm 0.45$ \\
\noalign{\smallskip}\hline
\end{tabular}
\]
\begin{list}{}{}
\item[$^\mathrm a$] The fluxes are given in units of
$10^{-17}\,\mathrm{erg}\,\mathrm{s}^{-1}\,\mathrm{cm}^{-2}$.  When
no uncertainty is given, the value is a 2$\sigma$ upper limit to
the flux in the line.
\item[$^\mathrm b$] These flux ratios are relative to $F(\mathrm
H\beta)$ measured for the CFHT spectrum on a scale where
$I(\mathrm H\beta) = 100$.
\end{list}
\end{table}

\begin{figure*}
\includegraphics[angle=90,width=18cm]{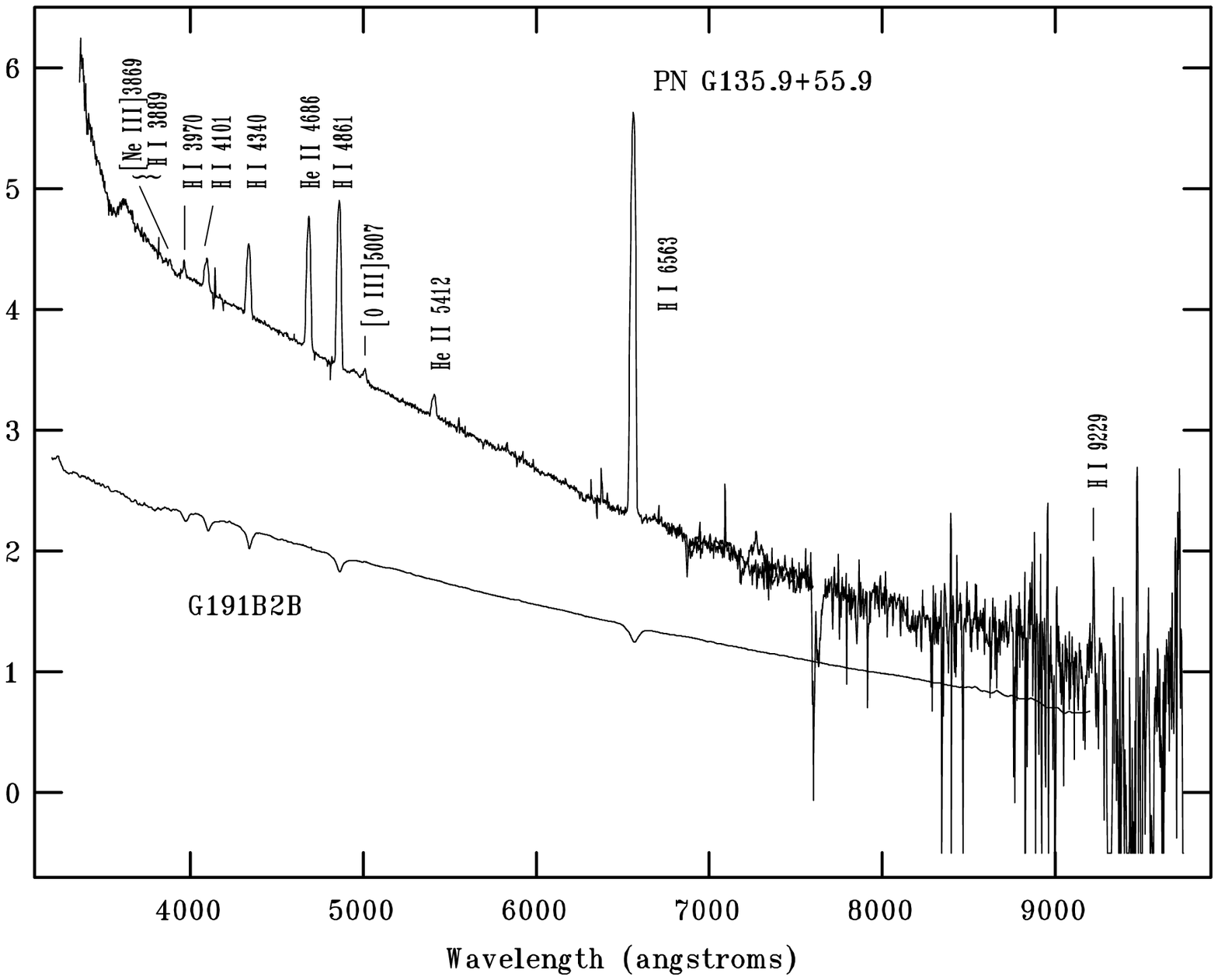}
\caption{We compare the spectra of \png\ and the standard star
G191B2B on an arbitrary magnitude scale.  For \png, we plot the
CFHT spectrum for $\lambda < 7500$\AA\ and the WHT spectrum for
$\lambda > 6760$\AA, neither corrected for reddening. The WHT
spectrum was normalized to the CFHT spectrum as described in the
text.  For G191B2B ($B-V = -0.32$\,mag), we plot the Oke
(\protect\cite{oke1990}) fluxes.  As noted by Tovmassian et al
(\protect\cite{tovmassianetal2001}), \png\ has a remarkably blue
continuum. } \label{pncontinuum}
\end{figure*}

Table \ref{whtlineint} presents the raw fluxes and the intensity
ratios relative to H$\beta$ for the WHT spectrum.  The only line
definitely detected is P9\,$\lambda$9229; P12\,$\lambda$8750 is
detected at only the $2\sigma$ level. Again, when no uncertainty
is given, the value represents a 2$\sigma$ upper limit to the line
intensity. These line intensities and limits were measured using
IRAF's splot routine. The fluxes represent the fluxes measured
directly in the summed WHT spectrum. The intensity ratios relative
to H$\beta$ were computed adopting the H$\beta$ flux from the CFHT
spectrum and correcting the WHT fluxes for the difference in the
slit widths used.  Based upon the spatial profile of H$\alpha$
from the CFHT spectrum, a 5\arcsec\ slit intercepts 3.32 times
more nebular emission than a 1\arcsec\ slit. The WHT fluxes were
then multiplied by this factor when computing the relative
intensities presented in Table \ref{whtlineint}.  Comparing the
continuum fluxes measured in the CFHT and WHT spectra in the
6760--7500\AA\ region, where the fringing in the CFHT spectrum is
not too severe, the continuum flux in the WHT spectrum should be
scaled upwards by a factor of 2.07 to match that in the CFHT
spectrum.  This scale factor is in good agreement with the value
of 1.95 expected based upon the 1\arcsec\ slit used for the WHT
spectra and the 1\farcs22 seeing measured from the spatial profile
of the continuum in the summed spectrum.

Generally, there is excellent agreement among the line intensities
over the wavelength range H$\delta$-\ion{He}{ii}$\lambda$5412.  In
the near-ultraviolet, the SPM spectrograph has very low efficiency
and the upper limits we derive from those spectra are considerably
less restrictive than the detections or limits from the CFHT
spectrum. We give upper limits to the [\ion{S}{ii}] line
intensities only for the last two SPM runs, since the CCD used for
the first two SPM observing runs and that at CFHT suffered from
fringing in the red.

\begin{table*}
\caption[]{Log of the direct imaging observations} \label{photlog}
\[
\begin{tabular}{lccccccc}
\hline \noalign{\smallskip} Telescope & Date           & CCD   &
Instrument & Filter$^\mathrm a$                & Exposure
time$^\mathrm b$ \\ \noalign{\smallskip}\hline\noalign{\smallskip}
SPM 2.1m  & 27-28 May 2001 & SITe3$^\mathrm c$ & Mexman     &
H$\alpha$\hfill (6565\AA, 11\AA)        & 3900s (5) \\
            &                &       &            & red cont.\ \hfill
(6650\AA, 46\AA)        & 1500s (3) \\ \noalign{\smallskip} NOT
2.6m  & 1-2 June 2001  & CCD7$^\mathrm d$  & ALFOSC     &
$y$\,\#18\hfill (5470\AA, 220\AA)          & 2700s (4) \\
            &                &       &            &
H$\alpha$\,\#21\hfill (6564\AA, 33\AA)        & 2700s (4) \\

\noalign{\smallskip}\hline
\end{tabular}
\]
\begin{list}{}{}
\item[$^\mathrm a$] The central wavelength and the bandpass
width for each filter are given in parentheses.
\item[$^\mathrm b$] The number of images is given in parentheses.
\item[$^{\mathrm c}$] This CCD has 24\,$\mu$m pixels in a
$1024\times 1024$ format.  Its gain and readnoise are 1.3\,e$^-
/\mathrm{pix}$ and 8\,e$^-$, respectively.  The plate scale is
$0\farcs312/\mathrm{pix}$.
\item[$^\mathrm d$] This CCD has 15\,$\mu$m pixels in a $2048\times
2048$ format.  Its gain and readnoise are $\sim 1\,\mathrm e^-
/\mathrm{pix}$ and 6\,e$^-$, respectively.  The plate scale is
$0\farcs188/\mathrm{pix}$.
\end{list}
\end{table*}

The notable exception to the good agreement among the line
intensities is H$\alpha$.  There is significant variation in the
${\mathrm H}\alpha/{\mathrm H}\beta$ ratio  among the summed
spectra for the different observing runs and between individual
spectra  for at least the SPM2 and SPM4 observing runs.
In both the SPM2 and SPM4 data sets, the dispersion among the
H$\alpha$ fluxes for the individual spectra also significantly
exceeds that for the H$\beta$ fluxes.  In the last line of Table
\ref{lineint}, we indicate the range of ${\mathrm
H}\alpha/{\mathrm H}\beta$ values found among the individual
spectra during each observing run. This variation is very
puzzling, since we normally obtained all of the spectra
consecutively on the same night (SPM2 is the exception). If this
variation is real, it is occurring (irregularly) on a time scale
of the order of an hour. Such behaviour is not at all expected in
a nebular plasma (e.g., Aller \cite{aller1987}). In a typical SPM
spectrum (of a half hour duration), at least 50,000, 15,000, and
2,500 photons are detected at H$\alpha$, H$\beta$, and H$\gamma$,
respectively, so the variation we see in the ${\mathrm
H}\alpha/{\mathrm H}\beta$ ratio would not appear to be due to
poor photon statistics. Over this wavelength range, we do not see
any variation exceeding more than a few percent in any of the
standard star observations. We made no effort to orient the slit
at the parallactic angle, but the wide slits used, particularly
for the 2001 observing runs, should compensate for the effects of
differential refraction. Regardless, were differential refraction
the culprit, we should see nearly equally large variations in
${\mathrm H}\gamma/{\mathrm H}\beta$ as we see in ${\mathrm
H}\alpha/{\mathrm H}\beta$ (Filippenko \cite{filippenko1982}), but
we see none.  Three instrumental effects, however, affect the 2001
data from SPM. First, the object was acquired by blind offset, so
the centering of the object in the slit was almost certainly not
optimal. Second, the offset guider is known to flex relative to
the instrument field of view, so the object centering was likely
somewhat variable for the 2001 observations at SPM. Finally, the
spectrograph was out of focus due the CCD being mis-aligned with
the camera's focal plane. It is not clear, however, how any of
these might introduce variations in the ${\mathrm
H}\alpha/{\mathrm H}\beta$ ratio alone without affecting other
line ratios.  None of these issues affect the CFHT data nor those
from SPM in 2002, yet the ${\mathrm H}\alpha/{\mathrm H}\beta$
variations exist in these data sets as well. Tovmassian et al.
(\cite{tovmassianetal2001}) found similar variations, from a
variety of observing sites, though they attributed them to the
poor observing conditions affecting their observations.  Although
unusual, it would appear that the variations in ${\mathrm
H}\alpha/{\mathrm H}\beta$ are real.

In Fig. \ref{pncontinuum}, we compare the spectra of \png\ and the
standard star G191B2B (Oke \cite{oke1990}). For \png, we plot the
CFHT spectrum for $\lambda < 7500$\AA\ and the WHT spectrum for
$\lambda > 6760$\AA, without applying any reddening correction to
either spectrum.  The WHT spectrum was scaled upwards by a factor
of 2.07, as described previously.  As noted by Tovmassian et al.
(\cite{tovmassianetal2001}), this planetary nebula has a
remarkably blue continuum.

Finally, these new data do not provide any direct diagnostic of
the physical conditions in the nebular plasma.  No density
diagnostic has been detected to date, though
[\ion{Ar}{iv}]$\lambda\lambda$4711,4740 might have been expected
given the high degree of ionization. Similarly, the only
temperature diagnostic available is the upper limit to the
[\ion{O}{iii}]$\lambda$4363/5007 ratio and it provides no useful
constraint unless the density is unusually high,
$10^{6}-10^{7}$\,cm$^{-3}$, which is excluded given the nebular
flux and size (see Sec. 4).

\subsection{Reddening}

From the Schlegel et al. (\cite{schlegeletal1998}) reddening maps,
the expected foreground reddening is $E(B-V)\sim 0.01-0.02$ mag.
The ${\mathrm H}\gamma/{\mathrm H}\beta$ and ${\mathrm
H}\delta/{\mathrm H}\beta$ ratios imply $E(B-V) \sim 0.3-0.35$
mag, based upon a temperature of $2\times 10^{4}$\,K, a density of
$10^{3}-10^{4}$\,cm$^{-3}$, the Storey \& Hummer
(\cite{storeyhummer1995}) line emissivities, and the Fitzpatrick
(\cite{fitzpatrick1999}) monochromatic reddening law parametrized
with a ratio of total-to-selective extinction of 3.041 (McCall \&
Armour \cite{mccallarmour2000}). On the other hand, the ${\mathrm
H}\alpha/{\mathrm H}\beta$ ratio implies a reddening $E(B-V) <
0.23$ mag for the same physical conditions, reddening law, and
line emissivities, even if we consider the largest line ratio we
observe, ${\mathrm H}\alpha/{\mathrm H}\beta = 3.5$. For ${\mathrm
H}\alpha/{\mathrm H}\beta$ ratios at the low end of the range
observed, the reddening is zero or negative.  The intensity of
$\mathrm P9/\mathrm H\beta$ from the WHT spectrum implies
$E(B-V)=0.05$\,mag.  We can also compute a reddening using
\ion{He}{ii}$\lambda\lambda$4686,5412. Adopting the same physical
conditions, reddening law, and line emissivities, we find negative
reddenings, i.e., \ion{He}{ii}$\lambda$4686 is too bright relative
to \ion{He}{ii}$\lambda$5412 by 12\%. In any case, it appears that
the reddening is at most modest, with $E(B-V) < 0.3-0.35$ mag.  In
the remainder of this paper, we shall assume that the reddening is
zero.  None of the conclusions would be affected had we adopted a
modest reddening.

\section{Imaging}

\subsection{Observations}

We obtained direct images of \png\, in narrow-band filters from
two sites (Table \ref{photlog}).  Images were obtained in the
emission line of H$\alpha$ and a nearby continuum bandpass with
the 2.1\,m telescope in SPM. The Mexman filter wheel was used for
observations. The seeing was modest, with stellar images having
FWHM of 1\farcs5 or worse.  Images were also obtained using ALFOSC
at the 2.6\,m Nordic Optical telescope (NOT) on La Palma, Canary
Islands, Spain. A narrow filter centred at H$\alpha$ was used, but
the continuum was measured using a Stromgren $y$ filter.  The NOT
images benefited from better seeing, $\approx1\farcs2$, on the
first night of this run.  The spectrophotometric standard stars
HZ44 and Feige 67 were observed along with the object on the
second night for calibration purposes. During both imaging runs,
sequences of at least 3 images in each filter were obtained, with
one long exposure usually not exceeding 20 min and two shorter
ones. This tactic allowed us to keep cosmic ray rate within
reasonable limits in order to later combine the images for
improved statistics and to permit the removal of cosmic rays.

The images were processed using IRAF.  The steps included bias
substraction, flat field correction, image combination, and
apperture photometry. The STSDAS package within IRAF was used for
the deconvolution of the images.

\begin{figure}
\centering \resizebox{\hsize}{!}{\includegraphics{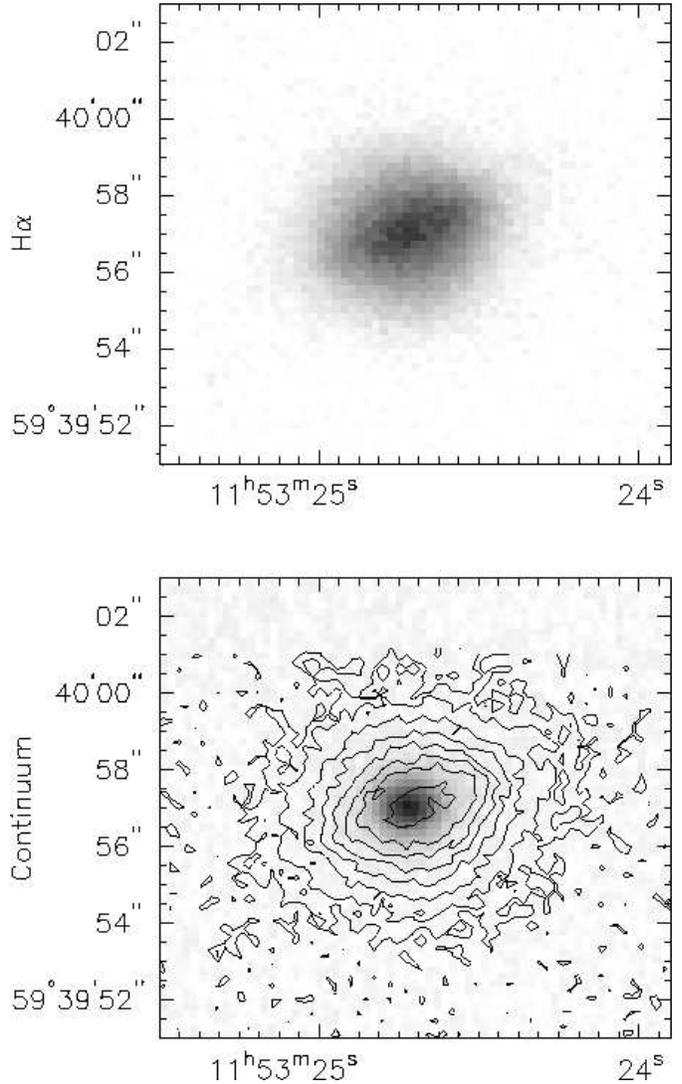}}
\caption{The upper
panel presents the image of \png\ in the light of H$\alpha$, while
the lower panel presents the Stromgren $y$ image of the central
star with the H$\alpha$ contours of the nebula superposed.  The
continuum and H$\alpha$ images were aligned using field stars.
Both images were obtained with the NOT.} \label{haimage}
\end{figure}

\begin{figure}
\centering \resizebox{\hsize}{!}{\includegraphics{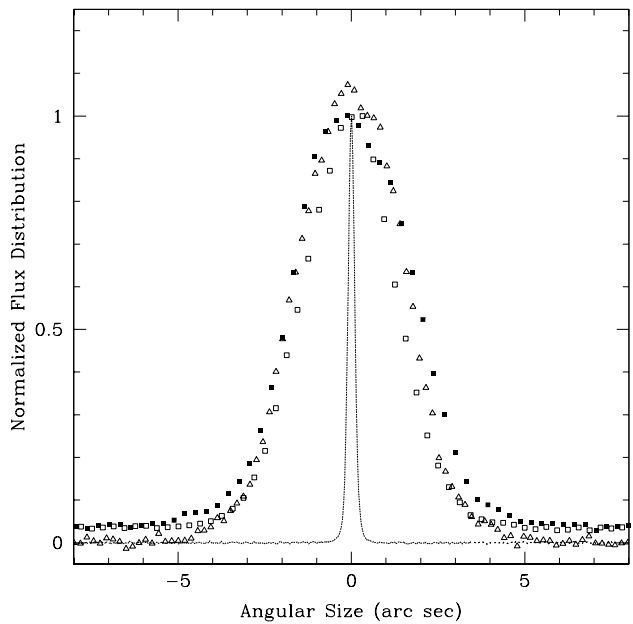}}
\caption{This figure presents the spatial profiles of the nebula
from the H$\alpha$ images and of the central object from the
6650\AA\ continuum image. The filled and open squares represent
cuts along major and minor axes of the nebula from the SPM images
\emph{after} subtraction of the continuum. Open triangles
represent the spatial profile along the major axis of the
H$\alpha$ image obtained at the NOT \emph{without} continuum
subtraction. The dotted line is the profile of the central star
(SPM). The fluxes are normalized to the maximum intensity of the
SPM continuum-subtracted image, except for the central star, which
is normalized to its maximum intensity.} \label{haprofile}
\end{figure}

\begin{figure}
\centering \resizebox{\hsize}{!}{\includegraphics{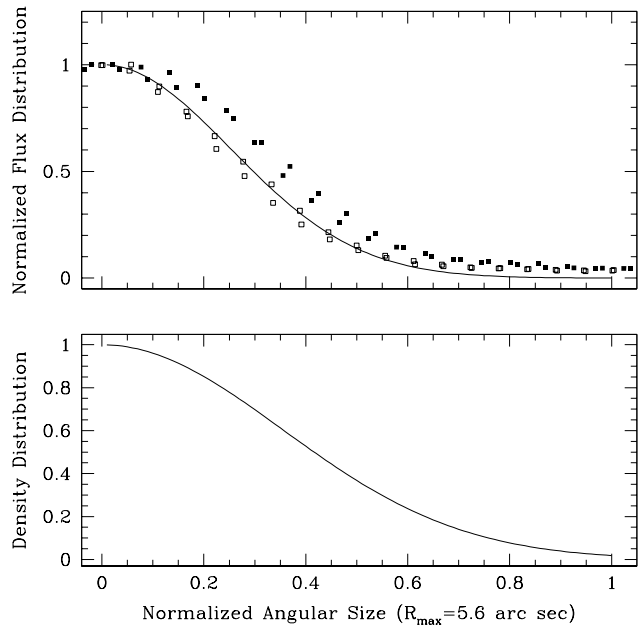}}
\caption{The density distribution used for the model calculations is
presented in the lower panel.  The predicted flux distribution
from our models is drawn with a solid line in the upper panel,
supposing that the emissivity is proportional to $n^2$ and
computed at a constant temperature.
The open and filled squares denote the spatial distributions of
the nebular emission along the major and minor axes, respectively,
obtained from the SPM observations, as shown in Fig \ref{haprofile}.
The abscissa denotes relative angular size normalized to the
maximum, azimuthally-averaged radius of 5\farcs6.  } \label{haprofile2}
\end{figure}

The primary uses of the H$\alpha$ images were to determine the
angular size of the nebula and to derive the density distribution
of the gas.  Two selected NOT images from the first night, where
the seeing was better and the focus was good were combined to
produce the image presented in Fig \ref{haimage}. The nebula
appears to be slightly elliptical in shape.  No substructure is
evident in the main body of the nebula. Its size comfortably
exceeds the PSF of stellar objects, so the instrumental profile
should not affect its shape.  Nonetheless, we conducted series of
deconvolution experiments on the nebular images using the
Lucy-Richardson  method (Lucy \cite{lucy1974}; Richardson
\cite{richardson1972}) that showed that the size and shape of the
nebula were unchanged after deconvolution.  This is our strongest
evidence that the central star does not emit significantly in
H$\alpha$.  For comparison the image of the central star obtained
in the Stromgren $y$ filter is presented in the bottom panel of
Fig. \ref{haimage} with the H$\alpha$ contours of the nebula
superposed. We used the SPM observations, where the continuum
filter was much closer in wavelength to the line filter, to
substract the image of central stellar object from that of the
nebula.  Figure \ref{haprofile} presents the spatial profiles of
the nebular emission in H$\alpha$ from the NOT image without
subtracting the central star and from the SPM image after
substraction of central star.  Except for a slight depression
around the peak, the shape of the nebula in both observations is
very similar.

The NOT observations in the Stromgren $y$ filter were used to
estimate the magnitude of the central star. We inferred
$m_{5556}=17.9\pm0.15$\,mag, corresponding to a flux of $F_{5556}
= 2.52\times 10^{-16}\,\mathrm{erg\,cm}^{-2}\,\mathrm s^{-1}\,\mathrm
\AA^{-1}$ in a good agreement with the flux determined from
spectroscopy.  This continuum measurement should be reliable since
there are almost no lines in the spectrum of this planetary
nebula, the contribution of \ion{He}{ii}\,$\lambda5413$\AA\ being
negligible.

\subsection{The nebular angular diameter and density distribution}

The angular diameter of the planetary nebula is an important
parameter since, in combination with the H$\beta$ flux, it allows
a determination of the electron density and the nebular mass. We
followed the recipe of Bedding \& Zijlstra
(\cite{beddingzijlstra1994}) to assure that our measurements
correspond to common standards.  (For cautionary remarks, see
van Hoof \cite{vanhoof2000}.)  There are a variety of definitions
of the diameter of a PN in the literature: the FWHM, the surface
brigthness contour at 10\% of the maximum surface brightness, and
the outermost contour at which emission is found. Since
PN~G\,135.9+55.9 appears to have a slightly elliptical shape in
the H$\alpha$ images, we have measured the diameter of the nebula
along two axes.  We find that the diameter of the nebula is
11\farcs4$\times$9\farcs8 if we use the outermost contour,
6\farcs3$\times$5\farcs2 if we use the 10\% surface brightness
contour, and 3\farcs5$\times$2\farcs95 if we use the FWHM.

Figure \ref{haprofile2} shows the density and flux distributions
adopted in our models.  The symbols correspond to the observed \Ha\
emission profiles across the major and minor axes obtained from the
SPM images (shown in Fig. \ref{haprofile}). The curve corresponds
to a spherical model with a density distribution described by $n =
n_{c} \exp -(\theta/2.8)^{2}$, where $\theta$ is the angular
distance to the center in seconds of arc. In this model, the
emissivity of \Ha\ is taken to be simply proportional to $n^{2}$
and is computed for a uniform temperature. Such a simple
representation accounts very satisfactorily for the observed \Ha\
emission profile.

In the analysis of the photoionization models that follows, we
adopt a nebular radius of 5\farcs6, i.e., similar to the diameter
of the outermost contour.

\begin{table*}
\caption[]{Some global properties of the sequences of models
investigated }
\label{properties}
\[
\begin{tabular}{cccccccc}
\hline \noalign{\smallskip}
\Tstar\ & $Q({\rm{H^{0}}})$ & $L(H\beta)$  & $d$ & $z$ & $R_{out}$ &
$M_{neb}$ &
$n_{c}$  \\

[K] & [erg s$^{-1}$] & [erg cm$^{-2}$ s$^{-1}$] & [kpc] &  [pc] &[cm]  &
[\Ms] & [cm$^{-3}$] \\
\noalign{\smallskip}\hline\noalign{\smallskip}
100\,000 & 6~10$^{47}$ & 0.48 & 25.0 & 20.7 & 1.8~10$^{18}$ & 0.78 & 125  \\
\medskip
100\,000 & 1~10$^{47}$ & 0.080 & 10.2 & 8.5 & 6.3~10$^{17}$ & 0.070  & 190 \\
125\,000 & 6~10$^{47}$ & 0.29 & 19.2 & 15.9 & 1.6~10$^{18}$ & 0.45 & 148  \\
\medskip
125\,000 & 1~10$^{47}$ & 0.048 & 7.8 & 6.5 & 6.1~10$^{17}$ & 0.044 & 223  \\
150\,000 & 6~10$^{47}$ & 0.20 & 16.0 & 13.3 & 1.3~10$^{18}$ & 0.29  & 163 \\
150\,000 & 1~10$^{47}$ & 0.032 & 6.5 & 5.4 & 4.8~10$^{17}$ & 0.027  & 245 \\
    \noalign{\smallskip}\hline
\end{tabular}
\]
\end{table*}

\section{Photoionization analysis: new limits on the oxygen abundance}

\subsection{The modelling procedure}

The observational data do not permit a direct estimate of the
oxygen abundance, since nothing is known of the electron
temperature or of the presence of oxygen ions more charged than
\Opp.  It is therefore necessary to rely upon photoionization
models.

We have constructed sequences of photoionization models in which
the oxygen abundance varies over several orders of magnitude. The
models are constrained by the available observations, which
consist of the observed line intensities, the equivalent width of
\Hb, the total flux in \Hb, the size of the nebula, and the radial
distribution of \Ha\ shown in Fig. \ref{haprofile}. As already
noted by Tovmassian et al. (\cite{tovmassianetal2001}), the shape
of the stellar continuum only implies that the star is hotter than
50,000~K. The models are computed with the photoionization code
PHOTO, using the atomic data listed in Stasi\'nska \& Leitherer
(1996). The central star is assumed to radiate as a blackbody of
temperature \Tstar. The hydrogen density at a radius $r$ is taken
to be $n$ = $n_{c}$ exp $-(r/h)^{2}$, where $n_{c}$ is a free
parameter and $h = 2.8 d / 2.06~10^{5}$, where $d$, the distance
to the nebula (in the same units as $h$ and $r$), is also a free
parameter. The ionizing radiation field is treated in the
outward-only approximation.  The computations start close to the
star and are stopped when the equivalent width in \Hb, W(\Hb),
becomes equal to the observed value, taken to be 70\AA.

With such a representation of the nebula, it is easy to show that,
for each assumed stellar temperature \Tstar\ and total luminosity
\Lstar, the total observed flux in \Hb\ implies a certain distance
to the planetary nebula. The observed angular size of the image
then fixes the outer radius of the nebula. By trial and error, we
determine the value of $n_{c}$ for which the radius $R_{out}$
corresponding to W(\Hb) = 70\AA\ is close to $ 5.6 d /
2.06~10^{5}$. Table \ref{properties} gives some global properties
of selected sequences of photoionization models.  This
includes the distance, $d$, and the height above the Galactic
plane, $z$, implied by these models. Note that $R_{out}$ and the
nebular mass, $M_{neb}$, are decreasing functions of the electron
temperature. The values given in the table correspond to the low
metallicity end of the model sequences.

The chemical abundances in each sequence of models are
parametrized by O/H, with the abundances of the heavy elements
with respect to oxygen following the recipe of McGaugh (1991).
 Note that the abundance ratios in \png, as in any PN, may be
significantly different from those assumed in McGaugh's recipe,
especially for carbon and nitrogen. For helium, we assume an
abundance of 0.08 with respect to hydrogen in all models. This is
close to the value estimated from the observed HeII 4686/\Hb\
ratio and using the upper limit for \ion{He}{i}\,$\lambda$5876
given by the CFHT spectrum (see Sect. 5).   In any event, our
estimate of the chemical composition of \png\ is independent of
the relative abundances adopted in the model sequences since our
estimate uses  the  observed line intensities and relies on
the ionization and temperature structure of the nebula, which
depend essentially on hydrogen and helium in the domain of
interest.

\begin{figure*}
\centering
\resizebox{\hsize}{!}{\includegraphics{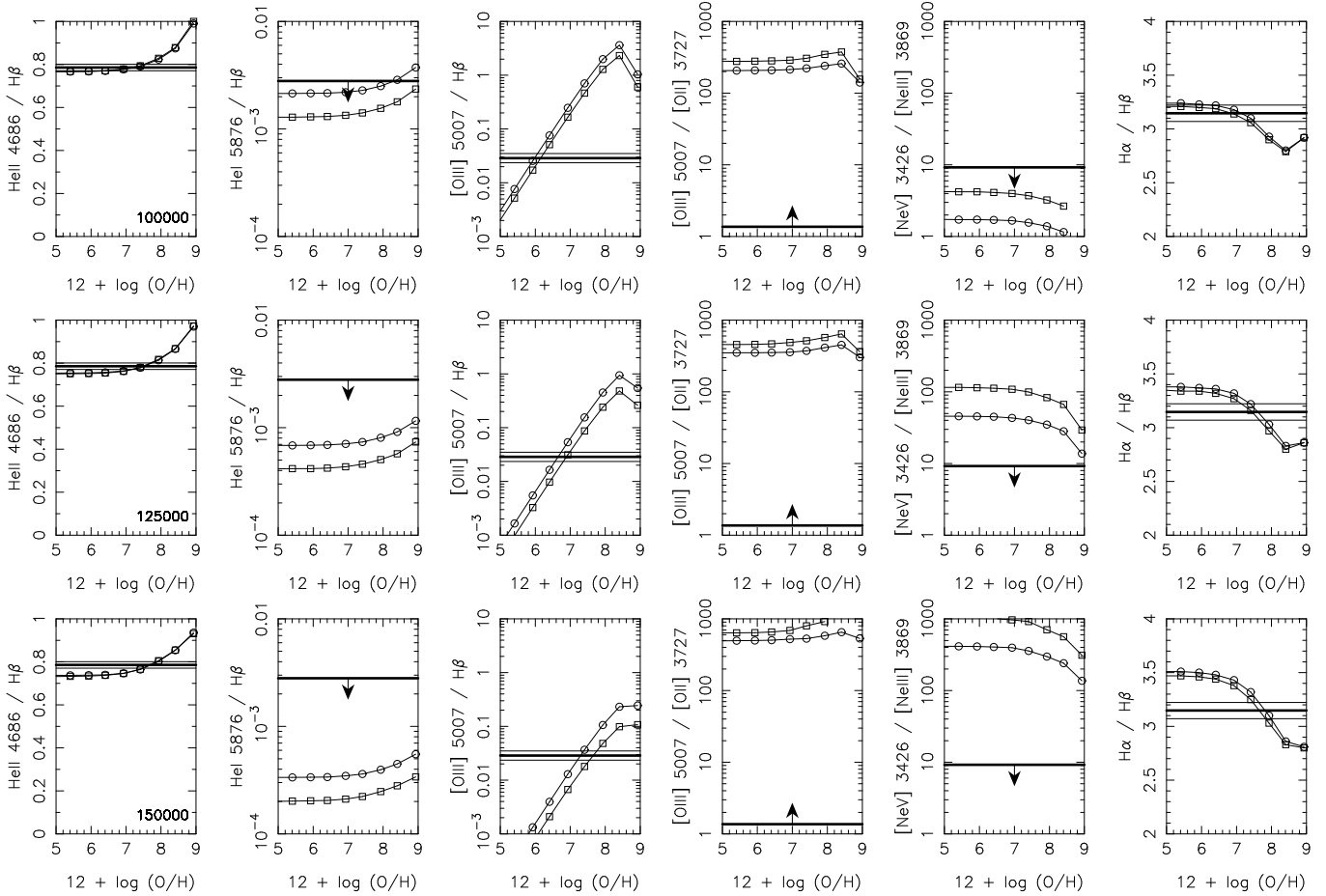}}
\caption{
This figure presents our grid of photoionization models.  Each row
of panels corresponds to a particular central star temperature, indicated
in the first panel.  Each column of panels plots a different line
intensity ratio.  In each panel, two
sequences of models are included: circles denote models with ionizing
luminosities of $10^{47}\,\mathrm{photons}\,\mathrm s^{-1}$, squares
denote models with ionizing luminosities of $6\times
10^{47}\,\mathrm{photons}\,\mathrm s^{-1}$.  The thick horizontal
lines indicate the observed values or their limits from the CFHT
spectrum.  In the case of limits, upward- and downward-pointing arrows
denote whether they are lower or upper limits, respectively.  Thin
horizontal lines denote the uncertainty range of measured values.}
\label{modelgrid}
\end{figure*}

\begin{figure*}
\centering \resizebox{\hsize}{!}{\includegraphics{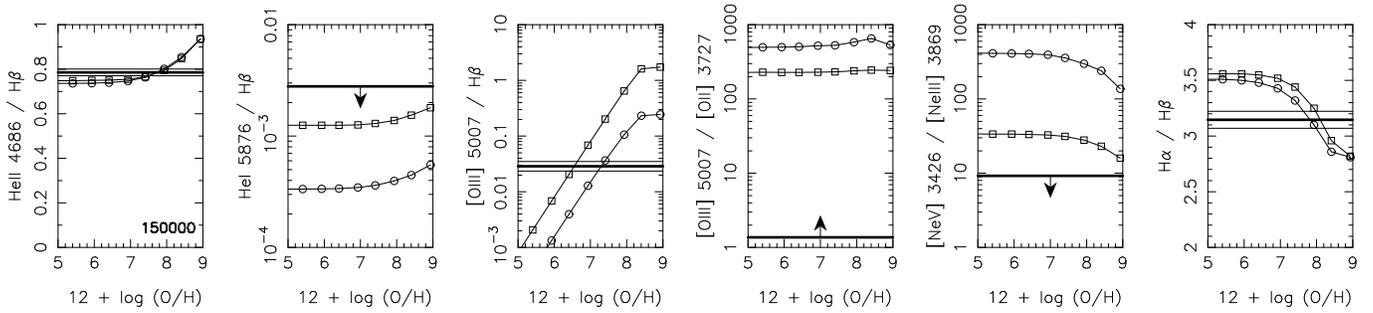}}
\caption{
Here, we compare two models with \Tstar\,$=150,000$\,K and
\qh\,$=10^{47}\mathrm{photon}\,\mathrm s^{-1}$, but with the circles
denoting a filling factor of 1 (the model from Fig. \ref{modelgrid})
and squares denoting a filling factor of 0.1.
As this experiment demonstrates, the maximum oxygen abundance allowed
by the models decreases as the filling factor decreases.}
\label{modelclumps}
\end{figure*}

Fig. \ref{modelgrid} presents the results of our models as a
function of $12 + \log (\mathrm O/\mathrm H)$.  Each row of panels
corresponds to a different central star temperature:
\Tstar\,$=100,000$\,K, \Tstar\,$=125,000$\,K, and
\Tstar\,$=150,000$\,K, as indicated in the first panel.  Each
column of panels displays a different line ratio. In each panel,
two series of models are represented with different symbols.
Circles correspond to models with central stars having a total
number of ionizing photons, \qh, equal to
$10^{47}\,\mathrm{photons\, sec}^{-1}$, while squares correspond
to models with a total number of ionizing photons of $6\times
10^{47}\,\mathrm{photons\, sec}^{-1}$, values that roughly bracket
the luminosities of the central stars of planetary nebulae as
computed by post-AGB evolutionary models (Bl\"ocker 1995).  The
horizontal lines show the observed values: thick lines for
measured values or limits, thin lines indicating the uncertainties
in measured values, and upward- and downward-pointing arrows
denoting lower and upper limits, respectively.  For the
observational data, we adopted the CFHT observations, since they
provide the most accurate measurements and the most stringent
limits. For the \Ha/\Hb\ ratio, however, we plot the value derived
from the SPM1 observations. As noted earlier, our observations
give inconsistent results for the \Ha/\Hb\ ratio. Values of
\Ha/\Hb\ around 3.1 are easily accounted for by our models, but
lower values are extremely difficult to explain. We will return to
this issue in Sect. 4.3.

\subsection{Discussion of the models}

The new observations allow us to eliminate the models with
\Tstar\,$\geq 125,000$\,K, since these produce a \nev/\neiii\
ratio much larger than is observed. Our models show that \Tstar\
should be around 100,000\,K.  \Tstar\ cannot be much lower than
this value since \Hei\ would then be observed.  Note that the
observed lower limit to \oiii/\oii\ provides no useful constraint
upon the ionization structure (or central star temperature) of our
models.

For the models with \Tstar\,$= 100,000$\,K shown in Fig.
\ref{modelgrid}, the oxygen abundances compatible with the
observed \oiii/\Hb\ ratio are $5.8 < 12 + \log \mathrm O/\mathrm H
< 6.3$\,dex. Allowing for a reasonable uncertainty in \Tstar\ implies
$12 + \log \mathrm O/\mathrm H \leq 6.5$\,dex.  The upper limit to
\nev/\neiii\ eliminates models excited by a star with a
temperature significantly above $100,000$\,K which would otherwise
permit $12 + \log \mathrm O/\mathrm H$ of the order of 7--8\,dex.

We have constructed other series of models to test the robustness
of these conclusions. For example, we have calculated models in
which the nebular radius, total nebular flux, and \Hb\ equivalent
width were varied, but the conclusions remain unchanged.

We have also considered models in which the gas is distributed in
small clumps with the same global density law as above, but with
an overall filling factor of 0.1.  Although the H$\alpha$ image is
smooth at our resolution, we cannot a priori exclude the presence
of small scale clumps or filaments. Clumpy models that fit the
observational constraints will result in  a lower global
ionization level than smoothly-distributed models. For the
purposes of illustration, Fig. \ref{modelclumps} compares the
smooth model of Fig. \ref{modelgrid} with \Tstar\,$= 150,000$\,K
and \qh\,$= 10^{47}\,\mathrm{ph\,sec}^{-1}$ (circles) with a
clumpy model whose filling factor is 0.1 (squares). Because the
ionization is lower in the clumpy model, the $\mathrm
O^{+++}/\mathrm O^{++}$ ratio is lowered with respect to the
smooth model, while the \Oiii/\Hb\ ratio is raised. Consequently,
the oxygen abundance compatible with the observed \Oiii/\Hb\ ratio
is \emph{smaller} than in the smooth case. Note that the model
presented here gives $12 + \log \mathrm O/\mathrm H \simeq
6.5$\,dex, but still violates the \Nev/\Neiii\ and \Ha/\Hb\
constraints. The total nebular mass of a model of given total \Hb\
luminosity and given radius is roughly proportional to the square
of the filling factor. Therefore, clumpy models that satisfy the
observational constraints will have lower nebular masses than the
corresponding models listed in Table \ref{properties}.  Finally,
if the density in the clumps were extremely high, models with high
metallicities, even as high as solar, could account for the weak
intensities of the forbidden lines, because these lines would be
quenched by collisional de-excitation. For this to occur,
densities exceeding $10^{6}\,\mathrm{cm}^{-3}$ are required for
both \Oiii\ and \Neiii. Given that the \Hb\ flux and equivalent
width as well as the size of the nebula are known, such high
densities would imply that the filling factor would have to be of
the order of $10^{-8}$ or less, which is highly unrealistic.

Our computations were made assuming that the star radiates as a
blackbody. However, a more realistic stellar atmosphere would give
a different spectral energy distribution for the ionizing photons,
particularly at the largest energies. One expects that extended,
metal-poor atmospheres could have larger fluxes at energies above
100\,eV.  This would increase the \nev/\neiii\ ratio and
consequently strenghten our conclusion that $12 + \log \mathrm
O/\mathrm H \le 6.5$\,dex. On the other hand, absorption by metals
could depress the number of photons able to produce \nev\ (Rauch
  2002), in which case a star with \Tstar\,$\sim 150,000$\,K
could become acceptable. In this case, however, the excitation of
the nebula would be lower than that predicted by blackbody models
with \Tstar\,$= 150,000$\,K and the line ratios would resemble
those produced by the blackbody model with \Tstar\,$=100,000$\,K,
again implying $12 + \log \mathrm O/\mathrm H \leq 6.5$\,dex. In
any case, the amount of metals in the atmosphere is not expected
to be large, unless the atmosphere contains dredged-up carbon. A
definitive answer to this problem can only come from a direct
measurement of lines from more highly charged ions.

Note that the distance we obtain for our object (see Table
\ref{properties}) indicates that it is located in the Galactic
halo, in agreement with its radial velocity (Tovmassian et al
\cite{tovmassianetal2001}). Its derived nebular mass is compatible
with the range of nebular masses derived for Magellanic Cloud PNe
(Barlow 1987).

To conclude this section, we emphasize that our new observational
data allow us to confirm that \png\ is an extremely oxygen-poor
planetary nebula, with an oxygen abundance less than 1/50 of the
solar value. Our models favour a value of $12 + \log \mathrm
O/\mathrm H$ between 5.8 and 6.5\,dex, compared with the solar
value of $12 + \log \mathrm O/\mathrm H = 8.83$\,dex
(Grevesse \& Sauval \cite{grevessesauval1998}). Our modelling
experiments indicate that this conclusion is independent of
plausible changes in the properties assumed for the central star
or nebular envelope.

\subsection{The \Ha/\Hb\ problem}

Our preferred models with \Tstar\,$= 100,000$\,K and low
metallicities are compatible with the \Ha/\Hb\ ratio derived from
the SPM1 data, as well as with the highest values found in
individual spectra from the other SPM observations (Table
\ref{lineint}, Fig. \ref{modelgrid}). However, as discussed
earlier, we find apparently significant variations of the \Ha/\Hb\
ratio between the different observing runs and even among
individual spectra obtained during individual runs. The majority
of the individual spectra indicate that \Ha/\Hb\ is below 3, as do
the observations reported by Tovmassian et al
(\cite{tovmassianetal2001}).

Because of collisional excitation of the hydrogen lines, \Ha/\Hb\
cannot have the recombination value, but is expected to be larger.
Collisional excitation is unavoidable at electron temperatures
above $\sim 15,000$\,K as soon as there is a small fraction of
residual neutral hydrogen.  The amount by which \Ha/\Hb\ exceeds
the recombination value (2.75 at 20,000~K, 2.70 at 30,000~K, using
the case B coefficients of Storey \& Hummer 1995) depends upon
both the electron temperature, which is higher for higher values
of \Tstar, and on the amount of neutral hydrogen. For example, the
models with \Tstar\,$= 125,000$\,K predict a value of \Ha/\Hb\
around 3.4 at low metallicities. With our observational
constraints, there is not much room for a drastic reduction of the
proportion of neutral hydrogen in our models.  Indeed, for a given
\Tstar\ and chemical composition, the proportion of \Ho\ at each
point in the nebula is completely determined by (and roughly
proportional to) $n_{e}/(L_{V}/R^{2})$, where $n_{e}$ is the local
electron density, $L_{V}$ is the stellar luminosity in the $V$
band and $R$ is the distance of this point to the star.
$L_{V}/R^{2}$ is a distance-independent quantity that relates the
stellar flux in the $V$ band to the angular distance of this point
to the star; $n_{e}$ is determined by the density distribution law
obtained from our \Ha\ images and the value of $n_{c}$ is imposed by
the observational constraints on W(\Hb) and the total angular
radius, as explained in Sect. 4.1. Our models of \png\ with
\Tstar\ = 100,000\,K indicate that  the \Ha/\Hb\ ratio should not
be lower than 3 if the nebula is metal-poor. Values of \Tstar\
significantly below 100,000\,K could be consistent with some of the
observed values of
\Ha/\Hb\  but they are excluded by the failure to observe \Hei\ in
this object.

Does this mean that the object is not as oxygen-poor as inferred
above?  There are several arguments against a higher abundance. In
our \Tstar\,$= 100,000$\,K models, \Ha/\Hb\,$< 3$ implies $12 +
\log \mathrm O/\mathrm H > 7.5$\,dex, but this is clearly
incompatible with the observed value of \Oiii/\Hb, as seen in Fig.
\ref{modelgrid}. At the other extreme, for our \Tstar\,$=
150,000$\,K models, \Ha/\Hb\,$< 3$ implies $12 + \log \mathrm
O/\mathrm H> 8$\,dex, which, though marginally compatible with the
observed \Oiii/\Hb, predicts a value for the \Nev/\Neiii\ ratio
far larger than observed.

Nor is it likely that any possible variability of the \Ha/\Hb\
ratio affects our oxygen abundances. Supposing that this variation
is real and refers entirely to the nebular radiation, one would
then expect the \Oiii/\Hb\ ratio to vary as a consequence of
variable ionization or temperature conditions, but this is not
seen.  All of the line intensities apart from \Ha\ remain
remarkably constant, including the \Oiii\ line.  Therefore, we
believe that the \Ha/\Hb\ problem does not affect our conclusions
as regards the oxygen abundance in \png.

However, this \Ha/\Hb\ problem is extremely puzzling. One does not
expect this ratio to vary in nebular conditions for an extended
object. One possibility could be that \png\ harbours a compact
disk. Such a suggestion has been made for other planetary nebulae
based upon either morphological, spectroscopic, or variability
arguments (He 2-25: Corradi 1995; IC 4997: Miranda \& Torrelles
2000, Lee \& Hyung 2000; and M 2-9: Livio \& Soker 2001).
Accretion disks in close, interacting binary systems have the
particularity of both being variable (Warner \cite{warner1995})
and having \Ha/\Hb\ ratios much smaller than 3, sometimes
attaining values below unity (e.g., Williams 1995). If \png\
contained such an accretion disk and if this disk contributed to
the emission of the hydrogen and helium lines in the central part
of the nebula, this could explain both the variability of the
\Ha/\Hb\ ratio in \png\ and the values of 2.7 or lower observed in
some of our spectra. However, an accretion disk would be an
unresolved point source in our images, and our deconvolution
experiments found no significant contribution to the H$\alpha$
emission from the central source. Likewise, the lack of
Dopper-broadened emission lines (Tovmassian et al
\cite{tovmassianetal2001}) also argues against an accretion disk
as the origin of a significant fraction of the line emission
(Warner \cite{warner1995}).

\section{The abundances of the remaining elements}

Our high signal-to-noise spectra enabled us to measure the
intensity of the \Neiii\ line, and to derive upper limits to the
intensities of \Siii\ and \Ariii, which allow us to make some
inferences on the abundances of Ne, S and Ar.

\begin{figure*}
\centering \resizebox{\hsize}{!}{\includegraphics{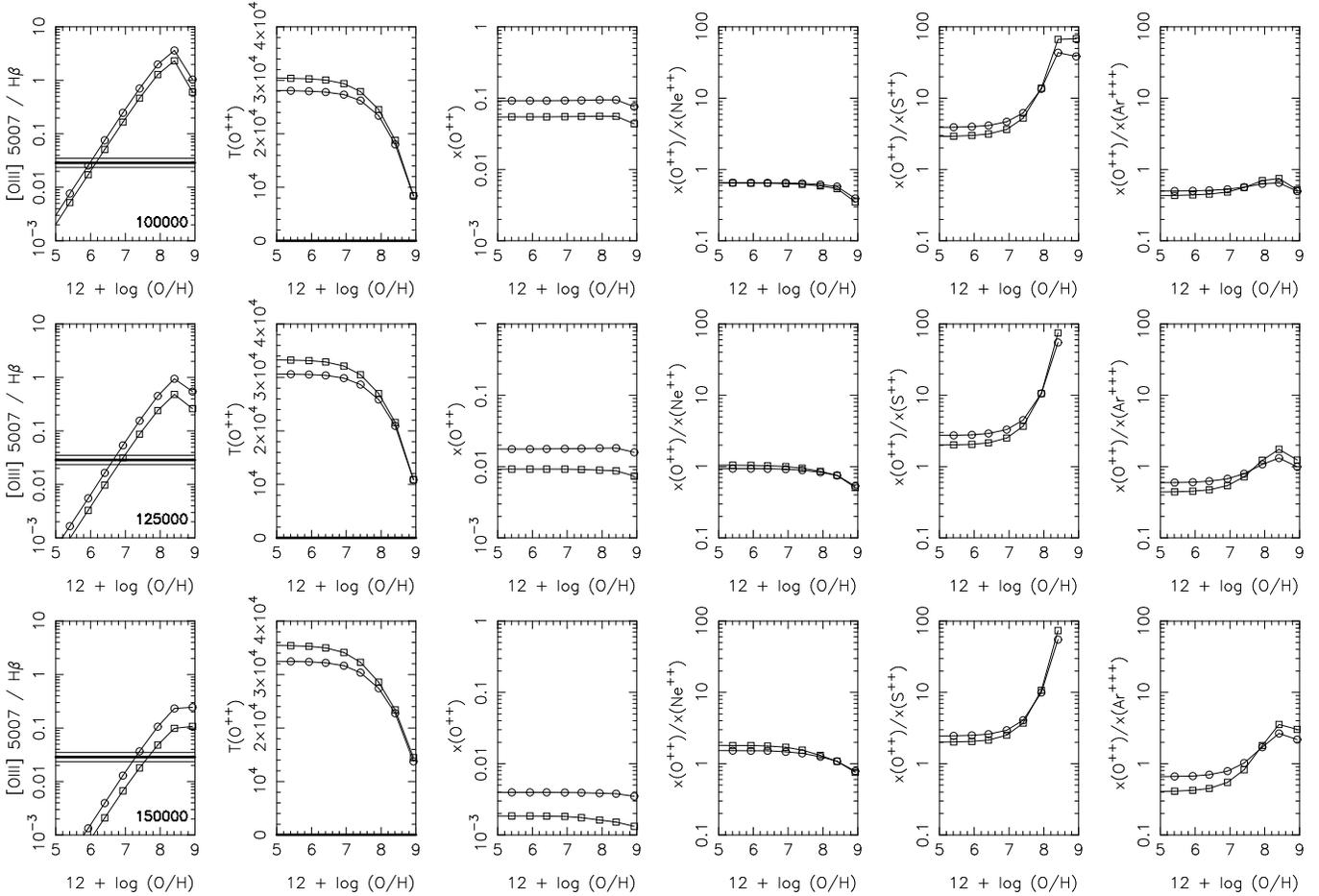}}
\caption{
This figure presents various physical parameters for the grid of
models in Fig. \ref{modelgrid}.  From left to right, the columns
present \Oiii/\Hb, the temperature in the O$^{++}$ zone, the weighted
fraction of oxygen in the form
of O$^{++}$, and the ionization correction factors for Ne$^{++}$,
S$^{++}$, and Ar$^{+++}$ with respect to O$^{++}$.  These data are
required to derive the abundances of Ne, S, and Ar in \png.}
\label{modelicfs}
\end{figure*}

To compute abundances, we must rely upon the electron temperature
and ionization structure provided by our models.  The relevant
quantities are shown in Fig. \ref{modelicfs} for the models from
Fig. \ref{modelgrid}. The first column of panels on the left
repeats the \Oiii/\Hb\ ratio from Fig. \ref{modelicfs} for
reference. The second column of panels gives the value of $T({\rm
O^{++}})$ defined as:
\begin {equation}
T({\rm O^{++}})=
\frac{\int{T_{e} n({\rm O^{++}}) n_{e}{\rm d}V}}{\int {n({\rm O^{++}})
n_{e}{\rm d}V}}.
\end {equation}

The next columns of panels give the ionization fractions
$x(\mathrm O^{++})/x(\mathrm H^{+})$, $x(\mathrm
O^{++})/x(\mathrm{Ne}^{++})$, $x(\mathrm O^{++})/x(\mathrm
S^{++})$ and $x(\mathrm O^{++})/x(\mathrm{Ar}^{++})$, where the
notation $x(\mathrm O^{++})$ stands for:
\begin {equation}
x({\rm O^{++}})=
\frac{\int{ n({\rm O^{++}}) n_{e}{\rm d}V}}{\int {n({\rm O})
n_{e}{\rm d}V}}.
\end {equation}

The value of $x(\mathrm O^{++})/x(\mathrm H^{+})$ is shown here
simply to demonstrate that $\mathrm O^{++}$ is far from being the
dominant oxygen ion in the nebula, and that a  more robust
determination of the oxygen abundance would require observing
lines from higher ionization stages.

Using the same atomic data as in the photoionization code and
taking as a characteristic temperature of the emission of the
\Neiii\ and \Oiii\ lines a value of 30,000~K derived from our
preferred models (Fig. \ref{modelgrid}),  we find that \Nepp/\Opp
$= 0.70\pm 0.4$, where the uncertainty reflects the uncertainties
of the \Oiii\ and \Neiii\ line intensities.  We furthermore note
that both \Opp\ and \Nepp\ are minority ions (Fig.
\ref{modelicfs}). Generally, one adopts \Nepp/\Opp = Ne/O. In our
preferred models (those corresponding to \Tstar\ = 100,000~K), we
find from Fig. \ref{modelicfs} that the ionization correction
factor is rather around 0.7, thus leading to $\mathrm{Ne}/\mathrm
O = 0.5\pm 0.3$.

In a similar fashion, our observed limits on the intensities of
the [\ion{Ar}{iv}]$\lambda\lambda$4711,4740 lines allow us to
derive that \Arppp/\Opp\,$< 0.45$. From our models (Fig.
\ref{modelicfs}) we find that the ionization correction factor is
0.5, and therefore Ar/O\,$< 0.23$.

Given that the nebula is strongly density-bounded, the observed
limits on the \Nii\ and \Sii\ line intensities give only crude
limits on abundance ratios involving these elements. With the
ionization correction factors from our preferred model, we find
N/H $<2.3\times 10^{-5}$ and S/H $<7\times 10^{-5}$. Such limits
are not very useful, except to infer that the N/H ratio is at most
equal to that in the Orion nebula.

Our infrared spectra allow us to estimate upper limits to the
intensities of \Siii\ and \Ariii, which are respectively $8\times
10^{-4}$ and $6\times 10^{-4}$  of the intensity of \Hb.  These
upper limits imply that $\mathrm S^{++}/\mathrm O^{++}< 0.031$ and
$\mathrm{Ar}^{++}/\mathrm O^{++} < 0.012$.  In our preferred
photoionization models, the ionization correction factor to derive
Ar/O from $\mathrm{Ar}^{++}/\mathrm O^{++}$ is around 20, so that
the upper limit on \Ariii\ implies Ar/O $< 0.24$ (in agreement
with the upper limit derived from
[\ion{Ar}{iv}]$\lambda\lambda$4711,4740). Regarding sulfur, we
find that the ionization correction factor is around 3 for S/O
from our models (Fig. \ref{modelicfs}), implying that $\mathrm
S/\mathrm O < 0.094$. It must be noted that the atomic data
concerning the ionization structure of S and Ar are uncertain (see
e.g. Ferland et al. 1998). Our models, were computed without
including dielectronic recombination to low-lying levels for these
ions. It is likely that the real ionization fractions of S$^{++}$
and Ar$^{+++}$ are actually higher than predicted by the models,
giving smaller ionization correction factors and more stringent
limits.

\begin{table}
\caption[]{$\alpha$-element abundance ratio comparison with other objects }
\label{abuncomp}
\[
\begin{tabular}{lccc}
\hline \noalign{\smallskip}
object                            & Ne/O      & S/O       & Ar/O \\
\noalign{\smallskip}\hline\noalign{\smallskip}
PN\,G\,135.9+55.9 $^{\mathrm a}$  &$0.5\pm 0.3$& $<0.094$  & $<0.23$ \\
Galactic disk PNe  $^{\mathrm b}$      &  0.26     & 0.017     & 0.005
\\
Galactic halo PNe  $^{\mathrm c}$      &  0.13     & 0.016     & 0.0016
\\
Orion nebula $^{\mathrm d}$       &  0.18     & 0.03      & 0.014
\\
Sun  $^{\mathrm e}$               &  0.18     & 0.03      & 0.004
\\
    \noalign{\smallskip}\hline
\end{tabular}
\]

\begin{list}{}{}
\item[$^{\mathrm a}$] This work.
\item[$^{\mathrm b}$] Kingsburgh \& Barlow (1994).
\item[$^{\mathrm c}$]  Howard et al. (1997).
These authors also show that  the  abundance ratios
are more dispersed in halo PNe  than in disk PNe.
\item[$^{\mathrm d}$] Esteban et al. (1998).
\item[$^{\mathrm e}$] Grevesse \& Sauval (1998).
\end{list}
\end{table}

It is interesting to compare the ratios of Ne/O, S/O and Ar/O we
find for \png\ with those of other kinds of objects. Table
\ref{abuncomp} shows the values for a sample of PNe in the
Galactic disk (Kingsburgh \& Barlow 1994)   and in the Galactic
halo (Howard et al. 1997),
for the Orion nebula
(Esteban et al. 1998), and for the Sun (Grevesse \& Sauval 1998).
This table is of course subject to uncertainties. Even the solar
abundances are quite uncertain for Ne, Ar and O (see Grevesse \&
Sauval \cite{grevessesauval1998}; the recent oxygen abundance
determination from Allende Prieto et al
\cite{allendeprietoetal2001} yields a value that is only 73\% that
obtained by Grevesse \& Sauval \cite{grevessesauval1998}). One
might then argue that the Ne/O ratio in \png\ is compatible with
the solar value, but we note that it is about twice the value
found in the Orion nebula and in disk planetary nebulae, where the
systematic errors in the abundance derivations are likely to be
similar. Supposing our Ne/O ratio is corect, it might indicate
some conversion of O into Ne by $\alpha$ capture. A few similar
cases are known among planetary nebulae (e.g., BB-1 has a
$\mathrm{Ne}/\mathrm O = 0.77$; Howard et al. 1997). Another
possibility is that the material from which the progenitor of
\png\ formed had an anomalous Ne/O ratio. Abundance studies of
metal-poor stars indicate that the very early galaxy was
chemically-inhomogeneous, with individual sites of star formation
being influenced by the explosions of nearby supernovae (e.g.,
Burris et al. \cite{burrisetal2000}).  The yields of O and Ne from
individual supernovae are also a function of the stellar mass
(Woosley \& Weaver \cite{woosleyweaver1995}; Thielemann et al
\cite{thielemannetal1996}), while observations of metal-poor halo
stars indicate that the scatter in oxygen abundances is
0.3-0.5\,dex at very low oxygen abundances (e.g., Israelian et al
\cite{israelianetal2001}).  As a result, it is probably not
surprising that the progenitor of \png\ might not have formed out
of material with the same Ne/O ratio that has characterized the
more recent Galactic disk.  Regardless, any conversion of O into
Ne is at most modest.  Even were \emph{all} of the Ne formed by
nuclear processing from oxygen, the initial oxygen abundance would
have been only 0.18\,dex larger than our preferred values. Thus,
the extremely low oxygen abundance in \png\ is genuine and not due
to nuclear and mixing processes in the progenitor star. The oxygen
abundance in \png\ should consequently reflect the chemical
composition of the medium out of which the star was made. The
limits we obtain on S/O and Ar/O in our object are consistent with
this view (although the limits are not very stringent).

The accuracy of the helium line intensities achieved in the
present observations raised the hope of obtaining very accurate
helium abundances. Taking the intensities determined from our CFHT
observations and adopting a temperature of 30,000~K  as inferred
from our models
and case B
coefficients from Storey \& Hummer (\cite{storeyhummer1995})  we
find that \Hepp/\Hp $= (7.50 \pm 0.14)\times 10^{-2}$. The upper
limit to the intensity of \Hei\ in the same spectra gives an upper
limit to \Hep/\Hp\ of $0.24\times 10^{-2}$ when using the
emissivities from Benjamin et al. (1999) in the low density limit.
A proper determination of the uncertainty in the derived helium
abundance should, however, account for collisional excitation of
the lines, a possible small amount of reddening, possible
deviations from case B, possible underlying absorption, as well as
temperature gradients inside the nebula. All of this can only be
attempted once the \Ha/\Hb\ problem is solved.  It is therefore
premature to propose an accurate value for the helium abundance in
\png.

\section{Conclusions}

Our new, extensive observations of \png\ confirm its
unusual nature.  Spectroscopy covering the wavelength interval
3400-9700\AA\ reveals lines of only \ion{H}{i}, \ion{He}{ii},
[\ion{O}{iii}], and [\ion{Ne}{iii}].  Deep H$\alpha$ imaging was
used to derive the radial density distribution.

Using these data as constraints, we constructed a new set of
nebular models from which we derived the abundance of oxygen and
the Ne/O, S/O, and Ar/O abundance ratios.  We confirm the
extremely low value of the oxygen abundance, which we find to be
less than 1/50 of the solar value: our models favour a value of
$12 + \log \mathrm O/\mathrm H$  between 5.8 and 6.5\,dex. The
distance implied by these models places \png\ in the Milky Way
halo, in accordance with its radial velocity (Tovmassian et al
\cite{tovmassianetal2001}). The models also imply nebular masses
in the range expected.

For the $\alpha$-element ratios, we find $\mathrm{Ne}/\mathrm O =
0.5\pm 0.3$, $\mathrm S/\mathrm O < 0.094$, and
$\mathrm{Ar}/\mathrm O
< 0.23$. The Ne/O ratio may be somewhat higher than is commonly found
in planetary nebulae in the Milky Way disk (e.g., Henry
\cite{henry1989}; Kingsburgh \& Barlow
\cite{kingsburghbarlow1994}). One possibility is that the
progenitor of \png\ converted some of its O to Ne. It is also
possible that the anomalous Ne/O ratio is the result of discrete
chemical enrichment in the very early evolution of the galaxy
(e.g., Burris et al. \cite{burrisetal2000}). Regardless, of the
cause, any conversion of O to Ne has been modest and does not
affect our conclusion that \png\ is the progeny of an
intrinsically very oxygen-poor star.

An unusual characteristic of \png\ is its low H$\alpha/\mathrm
H\beta$ ratio, for which we find no clear explanation.  Despite
its low metallicity and the concomitant high electron temperature
that should result in collisionally excited Balmer lines of
\ion{H}{i}, \png\ has an $\mathrm H\alpha/\mathrm H\beta$ ratio
typically below 3. Furthermore, $\mathrm H\alpha/\mathrm H\beta$
appears to be variable between observing runs and even within a
single night. One possible explanation for both the low $\mathrm
H\alpha/\mathrm H\beta$ ratio and its variability is if \png\
contains an accretion disk, though the evidence is not convincing.
At any rate, this issue does not appear to affect our conclusions
regarding the chemical abundances.

We also measure a low $\mathrm{He}/\mathrm H$ ratio of $\sim0.08$.
This makes \png\ interesting as a probe of the pregalactic He
abundance.  However, the derivation of a very precise He abundance
will require the resolution of a number of outstanding issues,
including the $\mathrm H\alpha/\mathrm H\beta$ problem, the
foreground reddening, and the internal temperature structure.

\begin{acknowledgements}

MGR thanks Anabel Arrieta, Leonid Georgiev, Felipe Montalvo, and
Salvador Monrroy for their able assistance with the observations
at SPM.  GT, GS, MGR, and CV are grateful for the receipt of
Director's discretionary time at the CFHT.  We are grateful to the
WHT staff for the spectrum obtained through service time.
GS acknoledges useful discussions with  Y. Izotov, J. M. Hur\'{e} and
M. Mouchet.
PDD is
a PPARC-supported PDRA. MGR acknowledges financial support from
DGAPA project IN100799 and CONACyT project 37214-E. GS
acknowledges financial support from DGAPA project IN114601.  GT
and GS acknowledge financial support from CONACyT project 34521-E.

\end{acknowledgements}

\end{document}